\documentclass[a4,12pt]{article}

\usepackage[sort&compress,numbers]{natbib}
\usepackage[margin=2.5cm]{geometry}
\usepackage{amsmath}
\usepackage{amssymb}
\usepackage{graphicx}
\usepackage{hyperref}
\usepackage{doi}

\graphicspath{
  {./}
  {./figures/}
  {./figures/wall/}
  {./figures/cartoon/}
  {./figures/envelopesketch/}
  {./figures/demo/}
}

\title{Gravitational waves from a first order electroweak \\
  phase transition: a brief review}

\author{David
  J. Weir\thanks{\href{mailto:david.weir@helsinki.fi}{david.weir@helsinki.fi}}
  \\[0.5cm] Department of Physics and Helsinki Institute of Physics,
  \\ P.O. Box 64, FI-00014 University of Helsinki, Finland}

\date{\today}

\begin{document}
\begin{flushright}
HIP-2017-06/TH
\end{flushright}

\begingroup
\let\newpage\relax
\maketitle
\endgroup

\begin{abstract}
We review the production of gravitational waves by an electroweak
first order phase transition. The resulting signal is a good candidate
for detection at next-generation gravitational wave detectors, such as
LISA. Detection of such a source of gravitational waves could yield
information about physics beyond the Standard Model that is
complementary to that accessible to current and near-future collider
experiments. We summarise efforts to simulate and model the phase
transition and the resulting production of gravitational waves.
\end{abstract}

\section{Introduction}

The fields of particle physics and cosmology are increasingly
intertwined. The discovery of the Higgs boson at the LHC has filled
one of the largest gaps in the Standard Model, although we may have
to wait for the next generation of colliders to see any evidence of
further physics beyond the Standard Model in the electroweak
sector. Meanwhile we have directly detected gravitational waves for
the first time, from binary black hole mergers, and the space-based
gravitational wave detector LISA is scheduled to launch in slightly
over a decade from now~\cite{Audley:2017drz}. In addition to studying
astrophysical processes, LISA will look for evidence of cosmological
phase transitions~\cite{Caprini:2015zlo}.

Although the phase transition in the electroweak sector of the
Standard Model would have been a 
crossover~\cite{Kajantie:1996mn,Gurtler:1997hr,Csikor:1998eu}, many
extensions of the Standard Model would undergo phase transitions
capable of emitting significant amounts of gravitational
waves. Furthermore, the signal from such a phase transition --
assuming it happened up to or around the TeV scale -- would be
perfectly placed for detection by LISA.

\begin{figure} % [!h]
\centering\includegraphics{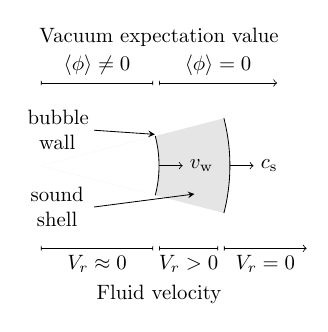}
\caption{Cartoon showing features associated with the bubble wall. The
  scenario shown is a subsonic deflagration, where the wall speed
  $v_\mathrm{w}$ is slower than the speed of sound $c_\mathrm{s}$. The
  scalar field bubble wall is shown, while the `sound shell' of
  nonzero fluid velocity in front of the wall is shaded. Above the
  diagram the value of $\langle \phi \rangle$ is shown, while below
  the radial fluid velocity $V_r$ is shown.}
\label{fig:cartoon}
\end{figure}

In this short review we summarise our current understanding of the
processes of gravitational wave production at a first-order phase
transition in the early universe. For the most part, we will
concentrate on the general case of a phase transition where bubbles of
the broken phase nucleate and expand in the presence of a plasma of
Standard Model particles. These particles exert a frictional force on
the wall, and a `sound shell' of plasma is excited in the vicinity of
the bubble wall (see Fig.~\ref{fig:cartoon}). We assume that the
frictional force is enough to stop the bubble wall from becoming
ultrarelativistic and ``running away''~\cite{Bodeker:2009qy}, which is
essentially always the case~\cite{Bodeker:2017cim}. However, there are
some phenomenological studies of gravitational wave production in
near-vacuum scenarios at higher energy
scales~\cite{Dev:2016feu,GarciaGarcia:2016xgv}, where there is no such
frictional force.

In the next section we will start by outlining in general terms the
electroweak phase transition and how it appears in several common
extensions of the Standard Model. This is followed in
Section~\ref{sec:energybudget} with a discussion of the motion of the
bubble wall and the resulting ``energy budget'' of the phase
transition. We summarise attempts to simulate and model bubble
collisions in Section~\ref{sec:simulations}, before attempting a
synthesis of the underlying gravitational wave production mechanisms
in Section~\ref{sec:processes}. We briefly show how to go from a
specific model to a predicted power spectrum in
Section~\ref{sec:frommodels} before looking towards future
developments in Section~\ref{sec:outlook}.

\section{The electroweak phase transition}
\label{sec:ewpt}

As discussed in the Introduction, without additional fields, the
electroweak phase transition is a crossover in the Standard Model,
occurring at a critical temperature of $159.5 \pm 1.5 \,
\mathrm{GeV}$~\cite{DOnofrio:2015gop}.

However, adding just a single extra scalar field -- real or complex;
whether a
singlet~\cite{Barger:2007im,Profumo:2007wc,Damgaard:2015con,Vaskonen:2016yiu,Beniwal:2017eik,Chen:2017qcz},
a second Higgs
doublet~\cite{Cline:1996mga,Fromme:2006cm,Dorsch:2013wja,Haarr:2016qzq}
or indeed a triplet (adjoint) Higgs
field~\cite{Gunion:1989ci,FileviezPerez:2008bj}, reopens the
possibility of a first-order phase transition at the electroweak
scale. Furthermore, these models all (to varying degrees) have regions
of parameter space that will not be excluded in the near future by
collider experiments~\cite{Curtin:2014jma}.

There are therefore two motivations to study gravitational wave
production from an electroweak phase transition.

First, and most importantly, it remains a well-motivated and
attractive possibility to produce the observed baryon asymmetry
through baryogenesis~\cite{Kuzmin:1985mm,Shaposhnikov:1987tw} (see
Ref.~\cite{Morrissey:2012db} for a review). Electroweak baryogenesis
fulfils the Sakharov conditions~\cite{Sakharov:1967dj} in the
following manner:
\begin{enumerate}
\item \textbf{$C$ and $CP$ violation}: this occurs due to particles
  scattering off the bubble walls, producing asymmetries in front of
  the walls.
\item \textbf{Baryon number $B$ violation}: The $C$ and $CP$ violation
  means that sphaleron transitions in front of the wall are biased to
  produce more baryons than antibaryons.
\item \textbf{Out of equilibrium}: The bubble walls (and associated
  sound shells) disturb the symmetric-phase equilibrium state.
\end{enumerate}
Even though the Standard Model is a crossover, and hence does not
depart far from equilibrium, it is possible to achieve
these requirements in the extensions mentioned above.

Second, a first-order phase transition at the electroweak scale would
source gravitational waves that are potentially detectable by
LISA~\cite{Caprini:2015zlo} (see
Refs.~\cite{Grojean:2006bp,Leitao:2012tx}, and also parts of
Ref.~\cite{Cai:2017cbj} for other reviews).  This would give a
complementary probe of the particle physics at this energy scale,
which will be studied extensively at planned experiments such as the
Future Circular Collider~\cite{Curtin:2014jma,Contino:2016spe}.

However, these two motivations are somewhat in tension. The energy
density in gravitational waves produced by a phase transition is
generally an increasing function of the wall velocity $v_\mathrm{w}$,
so faster wall speeds are desirable. However, the process of
electroweak baryogenesis outlined above depends on the wall velocity
relative to the plasma in front of the wall being slower than the
speed of sound~\cite{No:2011fi}, usually very much slower to allow
particles to diffuse from the bubble wall (where $C$ and $CP$
violation occur) back into the plasma (where biased sphaleron
transitions occur)~\cite{Joyce:1994fu}. Other variants of electroweak
baryogenesis which allow for a fast detonation have been proposed, for
example due to symmetry restoration behind the bubble
wall~\cite{Caprini:2011uz}, but further investigations -- and perhaps
simulations -- of such scenarios would be beneficial.

For the remainder of this review, then, we concentrate on the signal
from gravitational waves for its own sake, rather than as a signature
of a process which generated the baryon asymmetry in the early
universe.

\section{Motion of the bubble wall and the ``energy budget''}
\label{sec:energybudget}

As described above, a thermal first-order phase transition proceeds by
the nucleation of bubbles of the scalar field $\phi$ which is driving
the transition; this is typically the Higgs field, although in models
with additional scalar fields this is not always the case. The bubble
nucleation rate at temperature $T$ is given by
\begin{equation}
  \Gamma(T) = A(T) e^{-S_3(T)/T}
\end{equation}
where $S_3$ is the three-dimensional bounce solution and $A(T)$ a
dynamical prefactor of order $T^4$~\cite{Enqvist:1991xw}. The inverse
duration of the phase transition $\beta$ relative to the Hubble rate
$H_*$ at the time of the transition is then
\begin{equation}
  \frac{\beta}{H_*} = \left. \left[T \frac{\mathrm{d}}{\mathrm{d}T}
    \left( \frac{S_3(T)}{T} \right) \right] \right|_{T=T_*}
\end{equation}
where $T_*$ is the transition temperature, which we will assume for
simplicity is close to the nucleation temperature $T_\mathrm{n}$. We
will also assume that the duration of the phase transition is short
enough that expansion can be neglected (i.e. $\beta/H_* \gtrsim 1$). The
typical bubble radius $R_*$ is~\cite{Enqvist:1991xw}
\begin{equation}
  \label{eq:bubbleradius}
  R_* = (8 \pi)^{\frac{1}{3}} \frac{v_\mathrm{w}}{\beta},
\end{equation}
where $v_\mathrm{w}$ is the wall velocity. To a first approximation,
$R_*$ sets the inverse wavenumber of the peak of the gravitational
wave power spectrum from a thermal first-order phase transition.

The scalar field has stress-energy tensor
\begin{equation}
  T_{\mu\nu}^\phi = \partial_\mu \phi \partial_\nu \phi - g_{\mu\nu}
  \left( \frac{1}{2} \partial_\rho \phi \partial^\rho \phi - V(\phi)
  \right)
\end{equation}
where $V(\phi)$ is the classical potential.

We treat this $\phi$ as a background field which interacts with all
the particle content of the theory: Higgs bosons, quarks, leptons and
gauge fields. These form a plasma and, employing distribution
functions $f_i(k)$ for each particle species $i$, one finds that the
equation of motion for $\phi$ including the interactions with the
plasma can be written
as~\cite{Liu:1992tn,Moore:1995si,Konstandin:2014zta}
\begin{equation}
  \square \phi + \frac{\partial V(\phi)}{\partial \phi} +
    \sum_i \frac{\mathrm{d} m_i^2}{\mathrm{d} \phi} \int \frac{\mathrm{d}^3 k}{(2\pi)^3 2 E_i}
    f_i(k),
  \label{eq:wallphiunrenorm}
\end{equation}
where $m_i$ is the effective mass of the $i$th particle species
(including all gauge bosons, pseudo-Goldstone modes and fermions) and
$E_i^2 = k^2 + m_i^2$ (see
Refs.~\cite{Megevand:2009gh,Kozaczuk:2015owa} for discussions of this
approach in extensions of the Standard Model).

As the nucleated bubbles of the scalar field expand, they interact
with the plasma. This excites the plasma and creates a
`sound shell' around the wall of plasma moving with nonzero outward
radial velocity. Generally, if the wall velocity is smaller than the
speed of sound, then this shell precedes the scalar field wall and the
process is termed a `deflagration' by analogy with standard terms from
combustion physics. Conversely, if the wall velocity is faster than
the speed of sound, then the sound shell is a rarefaction wave
trailing the bubble wall and the resulting process is a `detonation'.

One can rewrite the equation of motion for the scalar field
\begin{equation}
  \square \phi + \frac{\partial V_\text{eff}(\phi, T)}{\partial \phi}
  = \mathcal{K}(\phi); \qquad \mathcal{K}(\phi) = - \sum_i \frac{\mathrm{d}
    m_i^2}{\mathrm{d} \phi} \int \frac{\mathrm{d}^3 k}{(2\pi)^3 2 E_i} \delta f_i(k),
  \label{eq:wallphi}
  \end{equation}
where $V_\text{eff}$ is the thermal effective potential, and $\delta
f_i(p)$ is the deviation of the distribution function of the $i$th
particle species from equilibrium.

\begin{figure} % [!h]
\centering\includegraphics{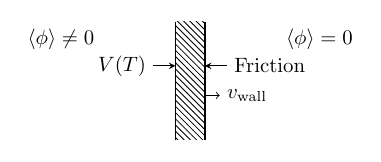}
\caption{Sketch of forces acting on bubble wall. The latent heat
  released during the phase transition drives the bubble outwards,
  while its interaction with the plasma of light particles creates
  friction. When the two forces are balanced, the wall ceases to
  accelerate.}
\label{fig:wallsketch}
\end{figure}

Equation~(\ref{eq:wallphi}) is important, for two reasons - firstly,
it underpins important simplifying approximations including the fluid
approximation that we shall use extensively throughout this work; and
secondly, it is readily apparent that the equation is nothing more than
a relationship between the outward force exerted by the bubble wall on
the particles $f_i(p)$, driven by latent heat, and the resulting
friction exerted on the bubble wall (see
Fig.~\ref{fig:wallsketch}). Nevertheless, the expression is difficult
to work with directly and so further simplifying assumptions are
usually made.

In particular, one often approximates the equilibrium distribution
functions for all the particle species $f_i$ by a relativistic fluid
$u^\mu$. The stress-energy tensor of such a fluid is
\begin{equation}
  T_{\mu\nu}^\text{fluid} = \sum_i \int \frac{\mathrm{d}^3 k}{(2\pi)^3 E_i}
  k_\mu k_\nu f_i(k) = w u_\mu u_\nu - g_{\mu\nu} p
\end{equation}
where $w = \epsilon + p$ is the enthalpy; $\epsilon$ is the energy
density of the fluid, and $p$ is the pressure.  Energy conservation
requires that the energy removed from the field $\phi$ by the friction
term $\mathcal{K}(\phi)$ is deposited in the fluid:
\begin{equation}
  \partial^\mu T_{\mu\nu} = \partial^\mu T_{\mu\nu}^\phi +
  \partial^\mu T_{\mu\nu}^\text{fluid} = 0.
\end{equation}

Working in the fluid approximation, one can take a more qualitative
form for $\mathcal{K}(\phi)$
\begin{equation}
  \mathcal{K}(\phi) = \eta(\phi, v_\mathrm{w}) u^\mu \partial_\mu \phi.
\end{equation}
The form of $\eta(\phi, v_\mathrm{w})$ is often chosen by comparison
with the Boltzmann equations for
$f_i(k)$~\cite{Huber:2013kj,Konstandin:2014zta}. Two choices that have
been used in numerical simulations are
\begin{equation}
  \eta(\phi,v_\mathrm{w}) = \text{const.} \quad \text{and} \quad
  \eta(\phi,v_\mathrm{w}) = \tilde{\eta} \frac{\phi^2}{T},
\end{equation}
where $\tilde{\eta}$ is a dimensionless constant. The exact choice of
$\eta(\phi,v_\mathrm{w})$ may slightly change the profile of the
scalar field and fluid at the bubble wall. However, as these are at
microscopic length scales when the phase transition occurs there is in
practice little difference. Furthermore, $\phi$ tends to a constant
and hence $\mathcal{K}(\phi) \to 0$ away from the bubble wall. We
therefore expect that the fluid sound shell reaches a scaling profile
parametrised by the dimensionless ratio $\xi = r/t$ and hence at
collision has a size proportional to $R_*$.

For the purposes of the gravitational wave power spectrum, then, the
scaling form of the radial fluid profile $V_r(\xi) [\equiv u_r/\gamma]$ and the wall velocity $v_w$
are all that matter. To know $V_r(\xi)$, one needs to know how much of
the latent heat ends up as fluid kinetic energy.

We first define the phase transition strength $\alpha$ as the ratio of
latent heat to radiation density at the time of transition in the
symmetric phase
\begin{equation}
  \alpha_T \equiv \frac{\mathcal{L}(T)}{g_* \pi^2 T^4/30},
\end{equation}
where $\mathcal{L}(T)$ is the latent heat and $g_*$ the number of
relativistic degrees of freedom at temperature $T$.  Note, however,
that another definition of $\alpha$ based on the trace anomaly
difference is sometimes used.

\begin{figure} % [!h]
\centering\includegraphics[width=4in]{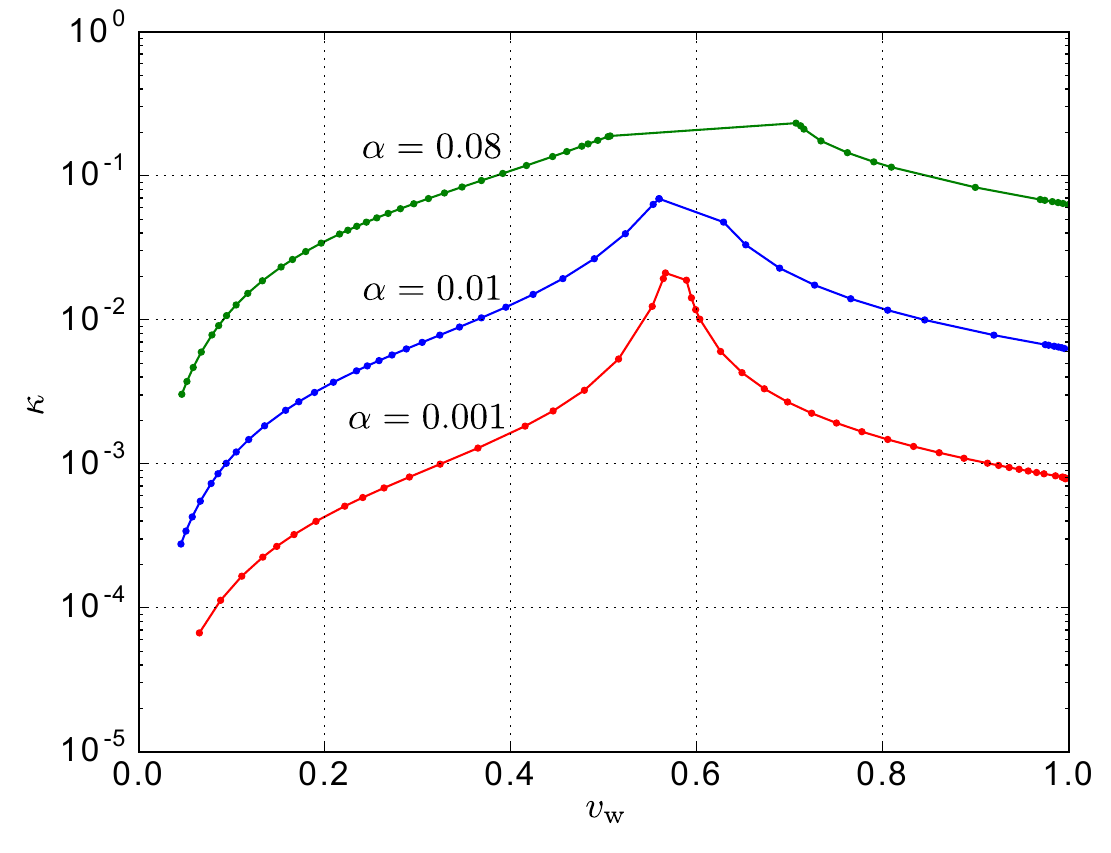}
\caption{Efficiency $\kappa_\mathrm{f}$ measured (at points marked by
  circles) from spherically symmetric simulations of the field-fluid
  system for a single bubble by means of
  Eq. (\ref{eq:efficiency})~(Cutting 2017, private
  communication). There is agreement with the analytically computed
  efficiency curves and in Ref.~\cite{Espinosa:2010hh}, even though
  the authors of that work used a bag model rather than the Standard
  Model-like effective potential employed here.}
\label{fig:efficiency}
\end{figure}

The fluid efficiency $\kappa_\text{f}$ then gives the fraction of this
vacuum energy that is turned into kinetic energy in the plasma during
the transition. It is
approximately~\cite{Espinosa:2010hh}
\begin{align}
  \kappa_\mathrm{f}(\alpha) \simeq \begin{cases}
    \dfrac{\alpha}{0.73 + 0.083 \sqrt{\alpha} + \alpha}, & v_\mathrm{w}
    \sim 1 \\[15pt]
    \dfrac{v_\mathrm{w}^{6/5} 6.9 \alpha}{1.36 - 0.037 \sqrt{\alpha} +
    \alpha}, &  v_\mathrm{w} \lesssim 0.1;
  \end{cases}
\end{align}
alternatively, if one knows the fluid velocity as a function of $\xi$
for a given scenario, the following expression can be used
\begin{equation}
  \label{eq:efficiency}
  \kappa_\text{f} = \frac{3}{\epsilon v_\mathrm{w}^3} \int \mathrm{d}\xi\, w(\xi) V_r^2
  \gamma^2 \xi^2.
\end{equation}
This expression has been used to produce the results shown in
Fig.~\ref{fig:efficiency}. The steady-state fluid equations of motion
can be solved to give the full profile for
$V_r(\xi)$~\cite{Espinosa:2010hh}, or it can be found from simulations
(see below).

For a given $\alpha_{T_*}$ and $v_\mathrm{w}$, there is essentially no
dependence on the microscopic details of the phase transition in
computing $\kappa_\mathrm{f}$, and there are relatively few parameters
required to adequately describe the physics of a thermal phase
transition: the inverse phase transition duration $\beta/H_*$, the phase
transition strength $\alpha_{T_*}$, and the wall velocity $v_\mathrm{w}$.

In the following section we show how these parameters can be used to
compute the gravitational wave power spectrum.

\section{Simulations, models and approximations}
\label{sec:simulations}

The first discussion of gravitational waves from a first-order
electroweak phase transition already anticipated a substantial
acoustic source~\cite{Hogan:1986qda}. Later works focused more on the
collision of the bubbles
themselves~\cite{Kosowsky:1991ua,Kosowsky:1992rz,Kosowsky:1992vn,Kamionkowski:1993fg},
and the `envelope approximation' -- infinitestimally thin walls that
disappear instantaneously when bubbles overlap -- gained wide
adoption.  High-precision studies were then carried
out~\cite{Huber:2008hg}.

Later it was observed that the fluid profiles are not infinitesimally
thin -- thus violating one requirement of the envelope
approximation -- and they do not disappear immediately after the
bubbles have collided, leading instead to an acoustic regime. Some
numerical work has also studied scalar field bubble
collisions~\cite{Kosowsky:1991ua,Child:2012qg}, also as a comparison
to the envelope approximation~\cite{Weir:2016tov}. However, it remains
that the envelope approximation and full realtime simulations with the
field-fluid model have been of the greatest interest. We discuss their
application to a general thermal phase transition below.

\subsection{Envelope approximation}

\begin{figure} % [!h]
\centering\includegraphics[width=2in]{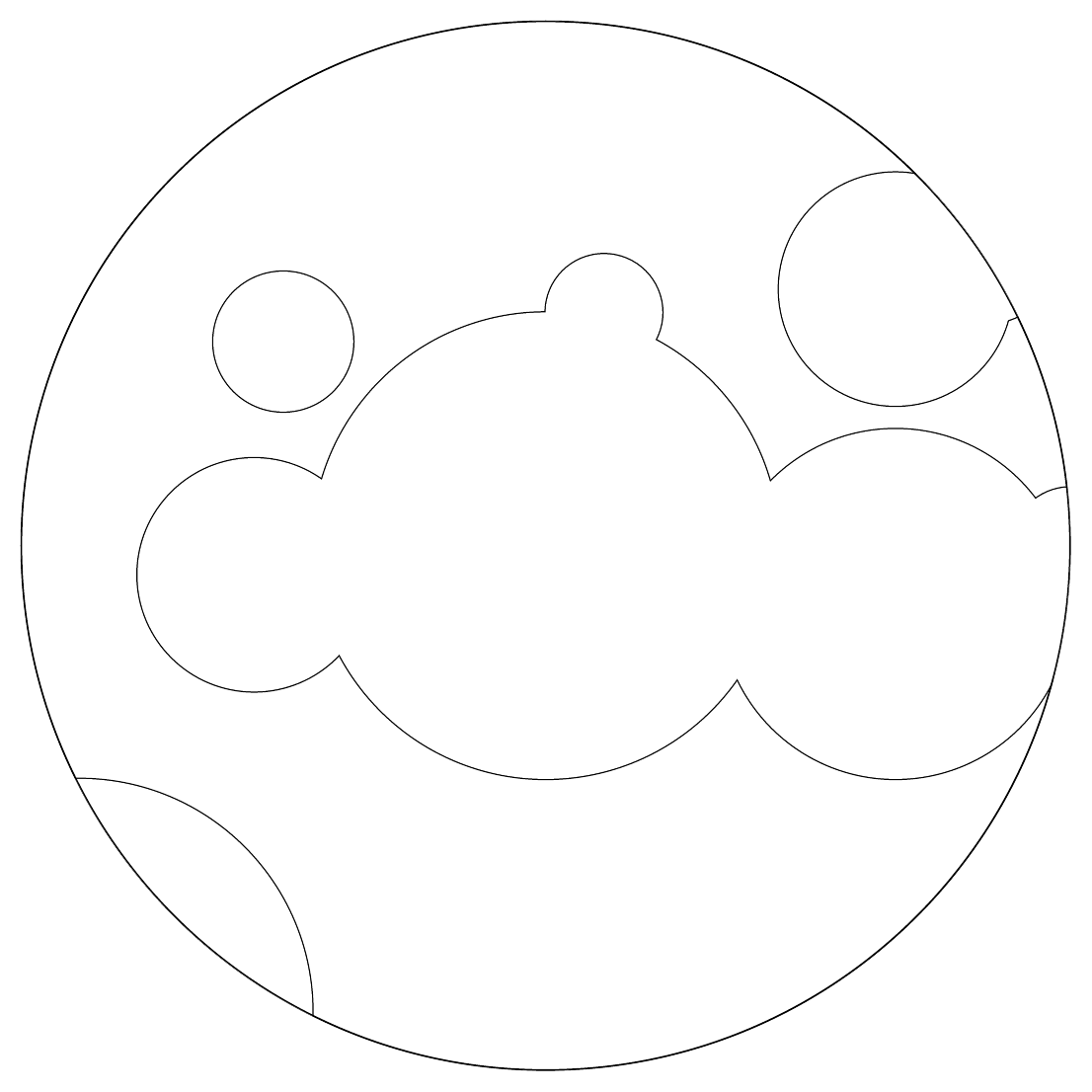}
\caption{Sketch of a slice through a `simulation' in the envelope
  approximation, with a spherical simulation volume. Only the
  uncollided portions of the thin bubble walls are recorded; there are
  no dynamics around the bubbles, or in the aftermath of bubble
  collisions.}
\label{fig:envelopesketch}
\end{figure}

The envelope approximation has been widely used in the past to model
gravitational wave power spectra from bubble collisions. It is really
two approximations: that the stress-energy
tensor of the expanding bubble is only nonzero in an infinitesimally
thin shell on the bubble's surface; and that this stress-energy
disappears immediately when two bubbles intersect, hence only the
`envelopes' of the bubbles interact (see Fig.~\ref{fig:envelopesketch}).

These two simplifying assumptions lead to a very simple power spectrum
-- a rising $f^3$ power law for frequencies much smaller than the
reciprocal bubble radius $1/R_*$, and a falling $f^{-1}$ for $f \gg
1/R_*$. This form has been confirmed by lattice simulations of
colliding scalar field walls~\cite{Weir:2016tov}, as well as
analytical modelling of coherent sums of infinitesimal fragments of
bubble wall~\cite{Jinno:2016vai}.

In Ref.~\cite{Huber:2008hg}, extensive studies of the form of the
gravitational wave power spectrum in the envelope approximation were
carried out. Based on their results, the authors postulated an ansatz
of the broken power-law form
\begin{equation}
  \Omega_\text{GW}(f) = \tilde{\Omega}_\text{GW} \frac{(a+b)
    \tilde{f}^b f^a}{b \tilde{f}^{(a+b)} + a f^{(a+b)}}
\end{equation}
where the power-law indices were (for fast walls) $a \approx 2.8$, $b
\approx 1.0$, $\tilde f$ is the peak frequency [a more complicated
  function of function of $\beta$ and $v_\mathrm{w}$ than the inverse
  of Eq.~(\ref{eq:bubbleradius})], and the amplitude
$\tilde{\Omega}_\text{GW}$ scales roughly as the cube of
$v_\mathrm{w}$.

In the past, the envelope approximation has been applied to all forms
of bubble collision, with the efficiency factor $\kappa$ taken to
refer to the efficiency of conversion of latent heat into fluid
kinetic energy, namely $\kappa_\mathrm{f}$. However, since the fluid
shells associated with the growing bubbles scale with the bubble
radius, it is not necessarily appropriate to make the approximation
that the bubble walls are infinitesimally thin. Furthermore, the
envelope approximation does not attempt to handle the aftermath of
bubble collisions.

For these reasons, the envelope approximation is best used for
modelling the scalar field contribution to first-order phase
transitions (which is only significant in certain circumstances), and
more sophisticated simulation and modelling techniques are required.

\subsection{The field-fluid model}

Motivated by the fluid approximation discussed in the previous
section, it is natural to consider both analytical and numerical
studies of the coupled field-fluid model. The equations of motion are
\begin{align}
(\partial_\mu \partial^\mu \phi) \partial^\nu \phi - \frac{\partial
    V_\text{eff}(\phi,T)}{\partial \phi} \partial^\nu \phi & = + \eta(\phi, v_\mathrm{w})
  u^\mu \partial_\mu \phi \partial^\nu \phi \label{eq:eomfield}
  \\
  \partial_\mu (w u^\mu u^\nu) - \partial^\nu p + \frac{\partial
    V_\text{eff}(\phi,T)}{\partial \phi} \partial^\nu \phi & = - \eta(\phi, v_\mathrm{w})
  u^\mu \partial_\mu \phi \partial^\nu \phi.
 \label{eq:eomfluid}
\end{align}

In a realtime numerical simulation of the system, the scalar field is
typically evolved using a standard leapfrog algorithm, while standard
operator-splitting grid-based techniques for the relativistic fluid
are required (see e.g. Ref.~\cite{WilsonMatthews}).

The microscopic physics of the sound shell, and the resulting
gravitational wave power spectrum, does not depend on the detailed
physics of the bubble wall. In simulations it is therefore usually
sufficient to consider a simplified effective potential
$V_\text{eff}(\phi,T)$ which yields the correct latent heat
$\mathcal{L}$.

It is relatively straightforward to solve the system of hydrodynamic
equations to find the scalar field and fluid velocity profile around
the bubble
wall~\cite{Kamionkowski:1993fg,Espinosa:2010hh,Huber:2013kj}, or else
one can evolve the above system of equations until a steady state is
reached.

When carrying out a full three-dimensional numerical simulation of the
system both the scalar field and the fluid source gravitational waves,
through the relevant transverse-traceless spatial parts of their
stress energy tensors,
\begin{equation}
  \tau^\phi_{ij} = \partial_i \phi \partial_j \phi; \qquad
  \tau^\mathrm{f}_{ij} = w u_i u_j.
\end{equation}
The largest three-dimensional lattice
simulations of the system performed to date use lattices with side
lengths of $4200$ sites. The smallest physically resolvable scales are
of the order of the spacing between sites, while the largest are
comparable to the size of the lattice itself. This means that there can
only be at most two or three orders of magnitude between the bubble
wall thickness and the bubble radius. Hence the
gravitational wave power sourced by $\tau_{ij}^\phi$ will be orders of
magnitude larger than it should be, relative to that sourced by
$\tau_{ij}^\mathrm{f}$. When extrapolating from the results of
numerical simulations, then, $\tau_{ij}^\phi$ is not included as a
source of gravitational waves.

\begin{figure} % [!h]
  \centering
  \includegraphics[width=2.8in]{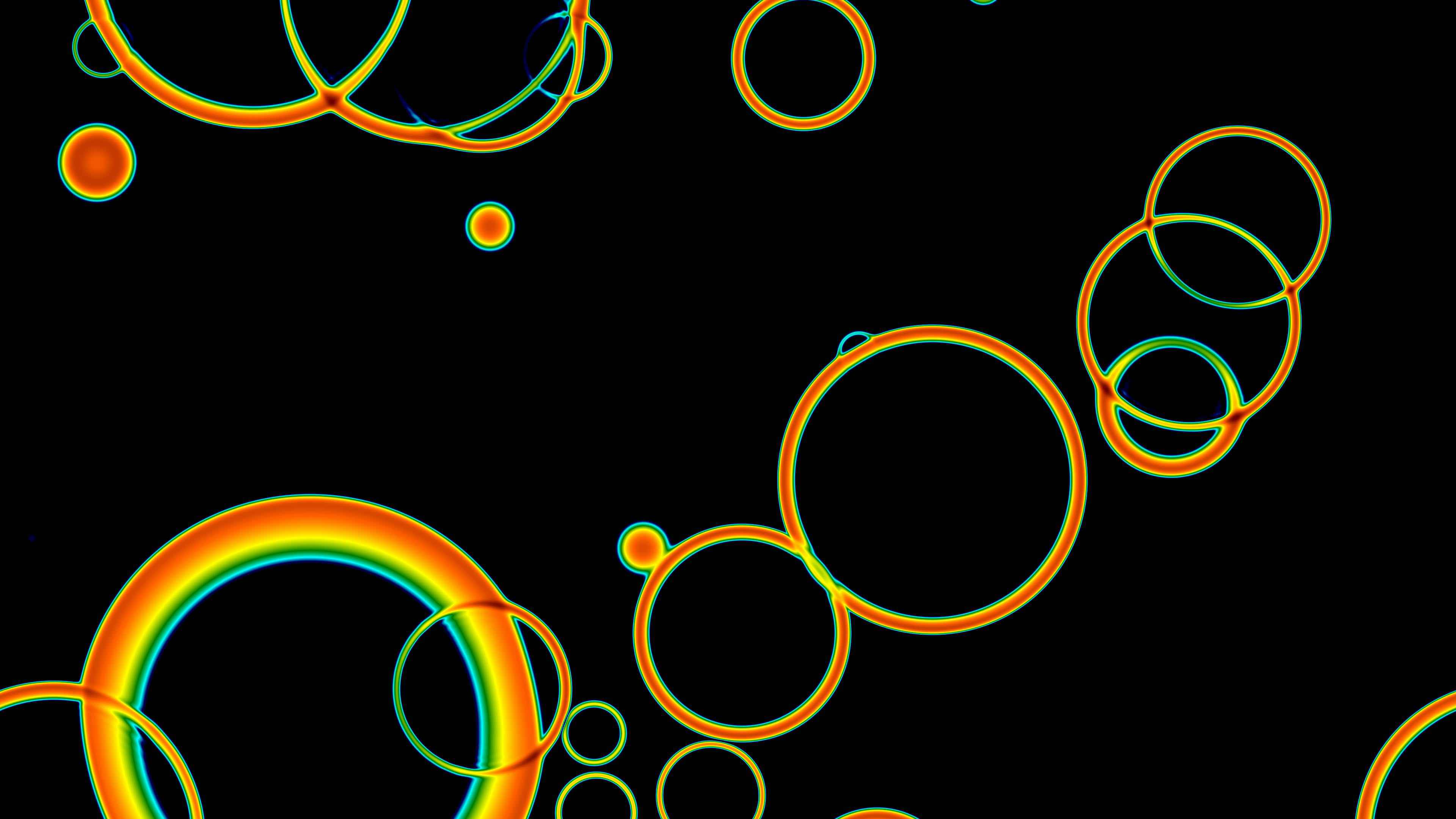}
  \includegraphics[width=2.8in]{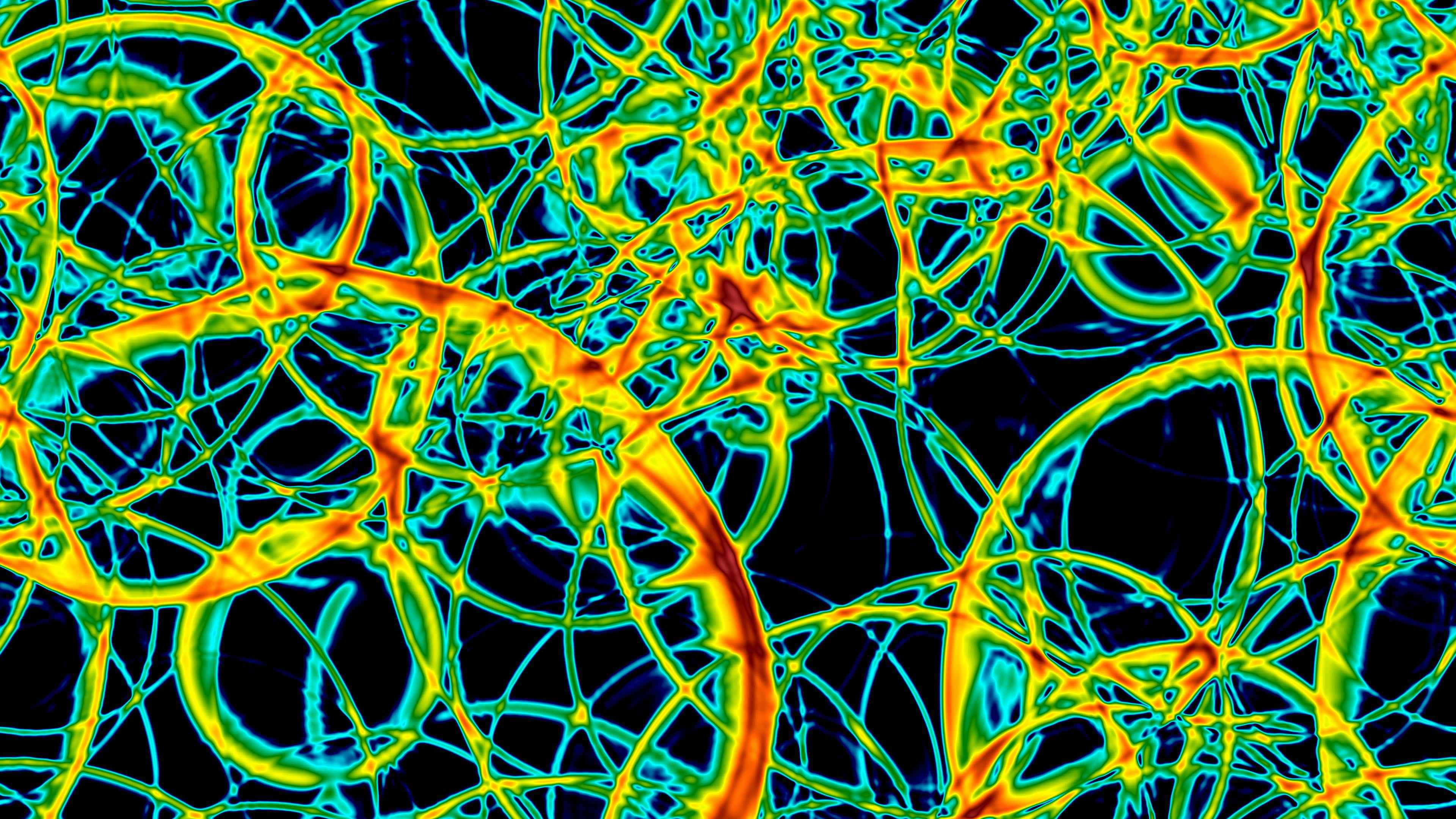}
\caption{Portions of slices through a three-dimensional field-fluid
  simulation, with hotter colours indicating relatively higher fluid
  kinetic energies. Here $\alpha_{T_*} \approx 0.01$ and $v_\mathrm{w}
  \approx 0.68$. The slice at left shows mostly uncollided bubbles,
  while the slice at right is from long after the bubbles have collided.}
\label{fig:slices}
\end{figure}

For further details about simulating the system of equations
(\ref{eq:eomfield}-\ref{eq:eomfluid}), see
Refs.~\cite{KurkiSuonio:1995vy,KurkiSuonio:1996rk,Giblin:2013kea}
(spherically symmetric simulations) and
Refs.~\cite{Hindmarsh:2013xza,Hindmarsh:2015qta,Giblin:2014qia,Hindmarsh:2017gnf}
(in three separate spatial dimensions). Portions of a slice through
some of the latest three-dimensional simulations are shown in
Fig.~\ref{fig:slices}.

\section{Gravitational wave production processes}
\label{sec:processes}

Based on the simulation results described in the previous section and
additional analytical calculations and modelling, we can now present
some ans\"atze for the resulting gravitational wave power spectrum. We
follow the discussion in Ref.~\cite{Caprini:2015zlo}, updated to
incorporate recent results~\cite{Hindmarsh:2017gnf}.

The production of gravitational waves at a first-order phase transition
can be separated into three stages.
\begin{itemize}
\item The first is the initial collision of the scalar field shells,
  which is of limited duration and generally subdominant unless the
  fluid efficiency is low or the system undergoes a vacuum transition
  in the absence of a thermal plasma. The gravitational wave power
  spectrum sourced by this stage is often denoted
  $\Omega_{\text{env}}$.
\item After the bubbles have merged, the wave of fluid kinetic energy
  in the plasma continues to propagate outwards into the broken
  phase. Without the driving force of the scalar field bubble wall,
  these waves travel at the speed of sound in the plasma. As the
  shells of kinetic energy from different bubbles overlap,
  gravitational waves are produced\footnote{Note that, for
    deflagrations, this sourcing of gravitational waves from
    overlapping sound shells may start before the scalar field walls
    collide, but as the source persists long after the initial
    collisions, we neglect this transient effect.}. The power spectrum
  produced by this source is denoted $\Omega_{\text{sw}}$.
\item Finally the acoustic phase may give way to
  shocks~\cite{Pen:2015qta} and a turbulent
  regime~\cite{Kamionkowski:1993fg,Kosowsky:2001xp,Nicolis:2003tg,Caprini:2009yp,Kahniashvili:2012vt}. The
  power spectrum is expected from analytical calculations to be rather
  different in this regime, but no simulations have yet captured time-
  and length-scales adequate to probe the onset of turbulence. We
  denote the resulting power spectrum $\Omega_{\text{turb}}$.
\end{itemize}

Peaking at different length scales, and on different time scales, the
three sources are expected to approximately sum together
\begin{equation}
  \Omega_\text{GW} = \Omega_{\text{env}} + \Omega_{\text{sw}} + \Omega_{\text{turb}}.
\end{equation}
Each source will contribute to a different extent, depending on the
exact details of the phase transition in question. For simplicity we
assume that the bubble wall does not run away, nor that it is
carefully tuned to produce a hybrid profile (with a wall velocity
close to the Chapman-Jouguet velocity). 

For the remainder of this section we summarise the form of these three
power spectra, motivated by simulations and analytic work. We will
consider ans\"atze for the amplitude of each of these sources at the
present day. For further information see Ref.~\cite{Caprini:2015zlo}.

\subsection{Colliding scalar field shells}

For the collision of scalar field shells the best available results
are those obtained from Refs.~\cite{Huber:2008hg,Jinno:2016vai}. Based
on the latter, we write the gravitational wave power spectrum as
\begin{equation}
  \label{eq:scalaransatz}
  h^2 \Omega_\text{env}(f) = 1.67 \times 10^{-5} \, \Delta
  \left(\frac{H_*}{\beta}\right)^2 \left( \frac{\kappa_\phi \alpha_{T_*}}{1 +
    \alpha_{T_*}} \right)^2 \left(\frac{100}{g_*} \right)^{\frac{1}{3}}
  S_\text{env}(f)
\end{equation}
with the spectral form (for $v_\mathrm{w}$ close to 1)
\begin{equation}
  S_\text{env}(f) = \left[ c_l
    \left(\frac{f}{f_\text{env}}\right)^{-3} + (1 - c_l -
    c_h)\left(\frac{f}{f_\text{env}} \right)^{-1} + c_h
    \left(\frac{f}{f_\text{env}}\right) \right]^{-1}
\end{equation}
where fitting yields $c_l = 0.064$ and $c_h = 0.48$ and the power law
indices are fixed. The peak frequency is
\begin{equation}
  \label{eq:peakfreqenv}
f_\text{env} = 16.5  \, \mu\mathrm{Hz} \,
\left(\frac{f_*}{\beta} \right) \left( \frac{\beta}{H_*} \right)
\left( \frac{T_*}{100 \, \mathrm{GeV}} \right) \left( \frac{g_*}{100}
\right)^{\frac{1}{6}}.
\end{equation}
The dependence of the amplitude and peak frequency on $v_\mathrm{w}$ is
\begin{equation}
\Delta =
 \frac{0.48  v_\mathrm{w}^3}{1 + 5.3 v_\mathrm{w}^2 + 5
  v_\mathrm{w}^4}; \quad
\frac{f_*}{\beta} = \frac{0.35}{1 + 0.069
  v_\mathrm{w} + 0.69 v_\mathrm{w}^4}.
\end{equation}

Into Eqs.~(\ref{eq:scalaransatz}) and~(\ref{eq:peakfreqenv}), one
inserts the transition temperature $T_*$, phase transition strength
$\alpha_{T_*}$, wall velocity $v_\mathrm{w}$ and nucleation rate
relative to the Hubble rate, $H_*/\beta$. Furthermore, the
`efficiency' factor $\kappa_\phi$ of converting vacuum energy into
scalar field gradient energy is required. This naturally depends on
both the surface tension and the surface area of bubbles at
collision. However, it is not straightforward to calculate the surface
area, which depends in a nontrivial way on the nucleation
rate~\cite{Weir:2016tov}. A very crude approximation would be
\begin{equation}
  \kappa_\phi \sim \frac{\gamma \sigma}{R_* \rho_\text{vac}},
\end{equation}
where $\gamma$ is the relativistic gamma associated with the wall
velocity, $\sigma$ is the surface tension, and $\rho_\text{vac}$ the
vacuum energy density. A more refined approach could be to use the
expression for the symmetric phase volume in
Ref.~\cite{Enqvist:1991xw} to infer the total surface area. For
general thermal phase transitions, which are the focus of this work,
we would expect $\kappa_\phi$ to be vanishingly small: as the walls
reach their terminal velocity, $\gamma$ approaches a constant, and so
the overall expression scales with $1/R_*$.

On the other hand, for runaway and vacuum transitions essentially all
of the vacuum energy goes into accelerating the bubble walls to
relativistic speeds. The efficiency factor $\kappa_\phi$ must then be
close to unity, and the gravitational waves are then principally
sourced by the scalar field gradient energy.

\subsection{Acoustic waves}

For a general thermal phase transition, the initial collisional phase
is short-lived; furthermore, the scalar field gradient energy scales
only as the surface area of the bubbles rather than the volume. A more
significant, and long-lasting source of gravitational waves is
produced by expanding sound shells in the fluid kinetic energy after
the bubbles have collided.

In fact, for non-ultrarelativistic fluid flows, it is straightforward
to obtain the gravitational wave power spectrum from acoustic waves
through a convolution of the fluid velocity
power~\cite{Caprini:2006jb}, and, in turn, this can be derived from a
fluid profile obtained through the methods discussed
earlier~\cite{Hindmarsh:2016lnk}. However, there is incomplete
agreement with the fluid velocity power spectrum observed in
simulations, perhaps due to the analytical work of
Ref.~\cite{Hindmarsh:2016lnk} not modelling the initial collisions of
the fluid profiles. We therefore concentrate for the time being on
results derived from recent very-large scale
simulations~\cite{Hindmarsh:2017gnf}.

The following ansatz for the gravitational wave power spectrum from
acoustic waves was first put forward in Ref.~\cite{Caprini:2015zlo}
and in Ref.~\cite{Hindmarsh:2017gnf} was found to generally agree with
simulation results. The version presented here is based on the latter work\footnote{Note that this equation contains several errors. Please see the \hyperref[sec:erratum]{Erratum} to this paper and to Ref.~\cite{Hindmarsh:2017gnf}.}:
\begin{equation}
  \label{eq:sw_power}
h^2 \Omega_\text{sw}(f) =
8.5 \times 10^{-6} \left(\frac{100}{g_*} \right)^{\frac{1}{3}}
\Gamma^2 \overline{U}_\mathrm{f}^4 \left(\frac{H_*}{\beta}\right)
v_\mathrm{w} \, S_\text{sw}(f)
\end{equation}
where the adiabatic index $\Gamma =
\overline{w}/\overline{\epsilon} \approx 4/3$; $\overline{w}$ and
$\overline{\epsilon}$ are the volume-averaged enthalpy and energy
density respectively.  The quantity $\overline{U}_\mathrm{f}$ is a
measure of the rms fluid velocity
\begin{equation}
  \overline{U}_\mathrm{f}^2 = \frac{1}{\overline w} \frac{1}{\mathcal{V}} \int_\mathcal{V}
  \mathrm{d}^3 x \, \tau_{ii}^\mathrm{f} \approx \frac{3}{4} \kappa_\mathrm{f} \alpha_{T_*}
\end{equation}
where the integral and average is over a volume $\mathcal{V}$. The
spectral shape is
\begin{equation}
   S_\text{sw}(f) = \left(\frac{f}{f_\mathrm{sw}}\right)^3  \left(
   \frac{7}{4 + 3 (f/f_\mathrm{sw})^2 } \right)^{7/2}
\end{equation}
with approximate peak frequency
\begin{equation}
  f_\mathrm{sw} = 8.9 \, \mu\mathrm{Hz} \, 
  \frac{1}{v_\mathrm{w}} \left( \frac{\beta}{H_*} \right) \left(
  \frac{z_\mathrm{p}}{10} \right) \left( \frac{T_*}{100 \,
    \mathrm{GeV}}\right) \left( \frac{g_*}{100} \right)^\frac{1}{6}
\end{equation}
with $z_\mathrm{p}$ a simulation-derived factor that is usually around
10, but may be higher when $v_\mathrm{w} \approx
c_\mathrm{s}$~\cite{Hindmarsh:2017gnf}.

We finish this section by making a comment on the timescale on which
shocks and then turbulence would
appear~\cite{LanLifFlu,Pen:2015qta}. It is given by the ratio
\begin{equation}
\label{eq:shocktime}
  \tau_\text{sh} \sim \mathrm{L}_f / \overline{U}_\mathrm{f},
\end{equation}
where $\mathrm{L_\mathrm{f}}$ is a measure of the characteristic length scale
associated with fluid flows -- to first approximation this is the
physical bubble radius $R_*$. Thus when the ratio $H_*
R_*/\overline{U}_\mathrm{f} \ll 1$, shocks can develop within a Hubble
time and the onset of turbulence must be taken into consideration.

\subsection{Turbulence}

Until simulations are available of the onset of turbulence, we must
make do with analytical results. From modelling of Kolmogorov-type
turbulence~\cite{Caprini:2009yp}, one obtains~\cite{Caprini:2015zlo}
\begin{equation}
  h^2 \Omega_\text{turb} (f) = 3.35 \times 10^{-4} \left(
  \frac{H_*}{\beta} \right) \left( \frac{\kappa_\text{turb} \alpha_{T_*}}{1
    + \alpha_{T_*}} \right)^{\frac{3}{2}} \left( \frac{100}{g_*}
  \right)^{\frac{1}{3}} v_\mathrm{w} S_\text{turb}(f).
\end{equation}
Here the quantity $\kappa_\text{turb}$ is the efficiency of conversion
of latent heat into turbulent flows. Based on simulation results so
far, at most a few percent of the fluid kinetic energy is converted
into rotational flow, so we might expect $\kappa_\text{turb}$ to be
negligible. However, we have not yet been able to study the timescale
of shock appearance [Eq.~(\ref{eq:shocktime})] in simulations, so it
remains likely that turbulent flows do form in many scenarios.

Although the amplitude is uncertain, the spectral shape of the
turbulent contribution is known exactly~\cite{Caprini:2009yp}
\begin{equation}
  S_\text{turb}(f) = \frac{(f/f_\text{turb})^3}{[1+(f/f_\text{turb})]^{\frac{11}{3}} ( 1
    + 8 \pi f/h_*) }
\end{equation}
where $h_*$ is the Hubble rate at $T_*$,
\begin{equation}
  h_* = 16.5 \, \mu\mathrm{Hz} \left( \frac{T_*}{100 \, \mathrm{GeV}
  }\right)  \left(\frac{g_*}{100} \right)^{\frac{1}{6}}.
\end{equation}
The peak frequency $f_\text{turb}$ is slightly higher than for the
sound wave contribution,
\begin{equation}
  f_\text{turb} = 27 \, \mu\mathrm{Hz} \,
  \frac{1}{v_\mathrm{w}} \left( \frac{\beta}{H_*} \right)
  \left(\frac{T_*}{100 \, \mathrm{GeV}}  \right)
  \left(\frac{g_*}{100}\right)^{\frac{1}{6}}.
\end{equation}

\section{From models to power spectra}
\label{sec:frommodels}

We have now discussed the means by which the three contributions to
the gravitational wave power spectrum can be studied analytically,
simulated and modelled.

\begin{figure} % [!h]
\centering\includegraphics[width=4in]{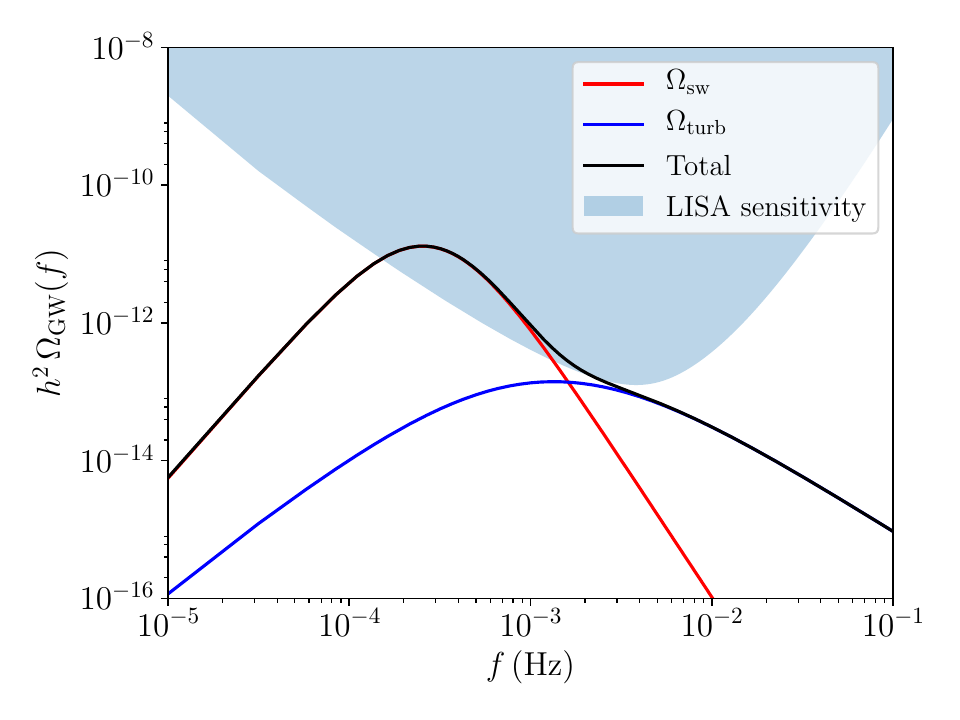}
\caption{Example of the gravitational wave power spectrum for a
  thermal phase transition, using the ans\"atze given in the text and
  with $v_\mathrm{w} = 0.44$, $\alpha_{T_*} = 0.084$, $H_*/\beta =
  0.1$ and $T_* = 180 \, \mathrm{GeV}$ (see
  Section~\ref{sec:frommodels}). The power spectrum is compared to a
  sensitivity curve obtained for a LISA-like configuration.}
\label{fig:demo}
\end{figure}

In Fig.~\ref{fig:demo}, we plot the gravitational wave power spectrum
based on the ans\"atze of the previous section, for a deflagration
with $v_\mathrm{w} = 0.44$, $\alpha_{T_*} = 0.084$, taking the
Standard Model value $g_* = 106.75$. Using the corresponding
simulation result from Ref.~\cite{Hindmarsh:2017gnf}, we find that
$z_\mathrm{p} = 6.9$, $\overline{U}_\mathrm{f} = 0.055$ and $\Gamma
\approx 4/3$. To turn these phase transition results into a possible
scenario, we use a transition temperature $T_* = 180 \, \mathrm{GeV}$
and take $H_*/\beta = 0.1$ (for which shocks are unlikely to develop
before Hubble expansion attenuates the signal).

We compare this example power spectrum with the sensitivity curve for
power laws (see Ref.~\cite{Thrane:2013oya}) for the eLISA
configuration closest to that proposed for LISA: 6 laser links, arm
length of $2~\mathrm{Gm}$ and mission duration of 5 years. In the
example given, the signal-to-noise ratio (SNR) should mean that
detection of such a scenario is possible. Nevertheless a careful
evaluation of the SNR is required~\cite{Thrane:2013oya,Caprini:2015zlo}.

To study the gravitational waves power spectrum resulting from a
specific extension of the Standard Model, one needs to supply at least
$\alpha_{T_*}$, $\beta$, $T_*$, and $v_\mathrm{w}$. This has been
done, for example, for the real singlet model in
Refs.~\cite{Vaskonen:2016yiu,Beniwal:2017eik}.

\section{Outlook}
\label{sec:outlook}

Gravitational waves produced by an electroweak phase transition are a
realistic candidate for detection by future space-based gravitational
wave detectors, such as LISA. The latest simulation and modelling
results indicate that it is principally the acoustic source that is
responsible for production of gravitational waves, although the role
of turbulence still requires clarification. The interplay between the
acoustic phase and the formation of shocks and turbulent behaviour is
still poorly understood. Further simulations are likely to be
required.

We are entering a period when the electroweak phase transition will
come under increasing scrutiny, in preparation for future colliders,
as well as for the detectability of gravitational waves. Precision
results for thermodynamic quantities in a wide variety of models are
required, possibly from simulations of dimensionally reduced models
(see e.g.~\cite{Brauner:2016fla} for the real singlet case). These
yield the phase diagram and hence $T_*$, but in addition, the latent
heat~\cite{Kajantie:1995kf} (and hence $\alpha_{T_*}$) as well as the
nucleation rate~\cite{Moore:2000jw} (and hence $\beta$) can be
determined. Combining these simulation results could yield a
computation of the gravitational wave power spectrum based almost
entirely on nonperturbative results. However, other techniques will
still be required to determine $v_\mathrm{w}$.

Throughout this paper we have specialised to the case of a bubble wall
where a terminal wall velocity $v_\mathrm{w} < 1$ is reached, rather
than a vacuum or runaway transition. Vacuum transitions have not been
studied extensively on the lattice. It is to be expected that the
envelope approximation performs well in these cases, however this
remains to be confirmed in future work.

Runaway transitions change the analysis slightly as they do not stir
up as much fluid kinetic energy, so the role of the colliding scalar
field bubble walls is likely to be more significant. However, since
higher-order corrections prevent true runaway transitions
from occurring~\cite{Bodeker:2017cim}, the analysis in this review
should be sufficient.

\bigskip

\noindent
\textbf{Competing Interests.} The author declares that they have no
competing interests. \\

\smallskip

\noindent
\textbf{Funding} I acknowledge PRACE for awarding me access to
resource HAZEL HEN based in Germany at the High Performance Computing
Center Stuttgart (HLRS). This work is supported by the Academy of
Finland grant 286769. \\

\smallskip

\noindent
\textbf{Acknowledgments} I acknowledge useful discussions with Mark
Hindmarsh and Kari Rummukainen. I am grateful to Daniel Cutting for
supplying Fig.~\ref{fig:efficiency}.

\bibliographystyle{apsrev4-1}

\bibliography{review_weir}

\newpage

\begin{center}
{\LARGE Erratum}
\end{center}

\setcounter{secnumdepth}{0}
\section[Erratum: Issues with the sound wave power spectrum formula]{Issues with the sound wave power spectrum formula}

\label{sec:erratum}

\noindent
Equation~(\ref{eq:sw_power}) based on Ref.~\cite{Hindmarsh:2017gnf} unfortunately includes a number of errors. Firstly, in Ref.~\cite{Hindmarsh:2017gnf}, the equivalent equation in the main body of the paper, Eq.~(45), appears without the reduced Hubble constant squared $h^2$ on the left hand side, where the present-day Hubble constant is defined as $H_0 = h \times 100 \, \mathrm{km}\;\mathrm{s}^{-1}\;\mathrm{Mpc}^{-1}$.

The remainder of the issues are corrected in the Erratum of Ref.~\cite{Hindmarsh:2017gnf}:
\begin{itemize}
  \item There was a factor of 3 missing from the right hand side of Eq.~(39), which yields a factor 3 in Eq.~(45).
  \item The numerical coefficient in Eq.~(45) should have been 0.687, not 0.68.
  \item The value quoted for $\tilde{\Omega}_\text{gw}$ in the main body was an order of magnitude too large and should be $\tilde{\Omega}_\text{gw}=1.2 \times 10^{-2}$.
\end{itemize}
Incorporating these changes, and multiplying both sides by $h^2$, with the contemporaneous Planck best-fit value $h=0.678$ (also used in Ref.~\cite{Hindmarsh:2017gnf}), we arrive at the following replacement formula for this paper's Eq.~(\ref{eq:sw_power}):
\begin{equation}
  h^2 \Omega_\text{sw}(f) =
  1.19 \times 10^{-6} \left(\frac{100}{g_*} \right)^{\frac{1}{3}}
  \Gamma^2 \overline{U}_\mathrm{f}^4 \left(\frac{H_*}{\beta}\right)
  v_\mathrm{w} \, S_\text{sw}(f),
\end{equation}
noting that $R_* = (8\pi)^{\frac{1}{3}}v_\mathrm{w}/\beta$ as in Eq.~(\ref{eq:bubbleradius}). The gravitational wave power spectrum shown in Fig.~\ref{fig:demo} is also therefore incorrect. An updated plot is, however, not included here for several reasons: our understanding of turbulence after a phase transition has improved since this paper came out; the expected LISA sensitivity curve has changed; and foregrounds are neglected.

As source modeling has improved since this paper was originally published, the interested reader is encouraged to compare the formulae in this paper with more recent results.

I apologise for any confusion caused.

\bigskip

\noindent
\textbf{Acknowledgments} I am grateful to Jenni H\"akkinen for a careful comparison of the sound wave power spectrum expressions found across the literature. I also acknowledge Pasquale Di Bari for pointing out that this review paper reproduced mistakes found in Ref.~\cite{Hindmarsh:2017gnf}, as well as introducing some more.

\end{document}